\documentclass[
superscriptaddress,nofootinbib,amsmath,amssymb]{revtex4-2}
\usepackage[colorlinks=true,linkcolor=blue,citecolor=magenta,linktocpage=true]{hyperref}
\usepackage{graphicx}% include figure files
\usepackage{bm}% bold math
\usepackage{hyperref}% add hypertext capabilities
\usepackage{mathtools}
\usepackage{soul}

\newcommand{\BH}{\texttt{BlackHawk} }
\renewcommand{\d}{{\rm d}}

\newcommand{\mrm}[1]{_{\rm #1}}

\makeatletter
\def\l@subsubsection#1#2{}
\makeatother

\begin{document}
\begin{flushright}

CERN-TH-2021-117

\end{flushright}

\title{Physics Beyond the Standard Model with BlackHawk v2.0}
\vspace{0.5cm}

\author{Alexandre Arbey}
\email{alexandre.arbey@ens-lyon.fr}
\affiliation{Univ Lyon, Univ Claude Bernard Lyon 1,\\ CNRS/IN2P3, IP2I Lyon, UMR 5822, F-69622, Villeurbanne, France}
\affiliation{Theoretical Physics Department, CERN, CH-1211 Geneva 23, Switzerland}
\affiliation{Institut Universitaire de France (IUF), 103 boulevard Saint-Michel, 75005 Paris, France}

\author{Jérémy Auffinger}
\email{j.auffinger@ipnl.in2p3.fr}
\affiliation{Univ Lyon, Univ Claude Bernard Lyon 1,\\ CNRS/IN2P3, IP2I Lyon, UMR 5822, F-69622, Villeurbanne, France}

\begin{abstract}
\vspace{0.5cm}
    We present the new version \texttt{v2.0} of the public code \texttt{BlackHawk} designed to compute the Hawking radiation of black holes, with both primary and hadronized spectra. This new version aims at opening an avenue toward physics beyond the Standard Model (BSM) in Hawking radiation. Several major additions have been made since version \texttt{v1.0}: dark matter/dark radiation emission, spin $3/2$ greybody factors, scripts for cosmological studies, BSM black hole metrics with their associated greybody factors and a careful treatment of the low energy showering of secondary particles; as well as bug corrections. We present, in each case, examples of the new capabilities of \texttt{BlackHawk}.
\end{abstract}

\maketitle

\tableofcontents

\section{Introduction}
\label{sec:introduction}

Since Hawking first demonstrated that black holes (BHs) evaporate by emitting a radiation close to the thermal radiation of a black body~\cite{Hawking:1974rv,Hawking:1974sw}, this phenomenon has been extensively studied. One of the main outcome of Hawking radiation (HR) is the possibility that small BHs (\textit{i.e.}~with mass $M\mrm{BH} \ll M_\odot$) formed just after the end of the universe expansion, namely primordial BHs (PBHs), may emit or have emitted radiation that could be observable today or that could have left an imprint in cosmology. This leads to a number of constraints on the PBH abundance, depending on their mass $M\mrm{BH}$, dimensionless spin $a^*$, or the extended distribution of both these parameters. For recent reviews on PBH formation and constraints, see Refs.~\cite{Carr:2020gox,Green:2020jor}. When considering PBHs with mass $M\mrm{BH} \gtrsim M\mrm{eva}$ where $M\mrm{eva} \simeq 5\times 10^{14}\,$g is the mass of (Schwarzschild) PBHs evaporating just today if formed at the beginning of the universe, these constraints are given as the fraction of dark matter (DM) $f\mrm{PBH}$ that the PBHs represent today. Indeed, light PBHs can represent some fraction of DM as they contribute to cold non baryonic matter in the early universe thermal history~\cite{Carr:2020xqk}. For lighter BHs with mass $M\mrm{BH} \lesssim M\mrm{eva}$, these constraints are given as the maximum possible fraction of the universe $\beta$ collapsed into PBHs at their formation. These BHs cannot represent a significant fraction of DM today since they have evaporated away. However, if evaporation stops at some point, leaving Planck scale remnants, these may contribute to DM~\cite{MacGibbon:1987my,Barrow:1992hq,Dolgov:2000ht,Chen:2002tu,Barrau:2003xp,Lehmann:2019zgt}. To determine the PBH evaporation constraints, one compares the Hawking radiation signals to astrophysical or cosmological observations such as \textit{e.g.}~the extragalactic gamma ray background (EGRB) or the $\Delta N\mrm{eff}$ limits respectively. PBHs in the mass range just above the evaporation limit, where energy is in the low energy QCD domain $E \sim$ MeV, are severely constrained as a fraction of DM. This represented a difficulty for \texttt{BlackHawk v1.2}, which relied on \texttt{PYTHIA} and \texttt{HERWIG} to compute the hadronization of primary particles, valid for $E\gtrsim$ GeV. There was a need for more careful treatment of this energy range, a work initiated by the study of Ref.~\cite{Coogan:2020tuf}. Noteworthy, HR is a purely gravitational phenomenon. As such, BHs would radiate any dark sector particle on top of the Standard Model (SM) spectrum, such as supersymmetric states~\cite{Baker:2021btk}, DM or dark radiation (DR)~\cite{Fujita:2014hha,Lennon:2017tqq,Hooper:2019gtx,Masina:2020xhk,Baldes:2020nuv,Hooper:2020evu,Keith:2020jww,Gondolo:2020uqv,Auffinger:2020afu,Masina:2021zpu,Arbey:2021ysg} --- and the graviton~\cite{Dolgov:2000ht,BisnovatyiKogan:2004bk,Anantua:2008am,Dolgov:2011cq,Dong:2015yjs,Inomata:2020lmk,Hooper:2020evu} (which contributes to DR).

It is thus of utmost importance to be able to predict precisely the HR spectra of SM and beyond SM (BSM) particles. A lot of analytical and numerical results exist in the literature for this purpose, the ones of~\cite{MacGibbon:1990zk,MacGibbon:1991tj,MacGibbon:1991vc,Halzen:1991uw} being among the most cited. However, people use most of the time the high-energy limit for the greybody factors given in these papers, resulting in order-of-magnitude constraints. There was thus a need for an automatic program to compute the precise spectra for any BH mass and spin, with a wide choice of parameters that allow for any kind of HR study. This was the context of creation of the public code \texttt{BlackHawk v1.0}~\cite{manual}. Since its first release, the code has received some modifications and has reached version \texttt{1.2}. The previous version of the manual can be found on the arXiv~\cite{manual} (v2), while the code is publicly available on HEPForge:
\begin{center}
    \url{https://blackhawk.hepforge.org/}
\end{center}
The code has been recently presented to the TOOLS 2020 conference~\cite{Auffinger:2020ztk}. \texttt{BlackHawk} is used by many groups from very different domains of astrophysics and cosmology to perform striking studies including, to the knowledge of the authors, evolution of BHs spin~\cite{Arbey:2019jmj}, EGRB constraints with extended mass distributions and spinning BHs~\cite{Arbey:2019vqx} or with higher dimensional Schwarzschild BHs~\cite{Johnson:2020tiw}, electron and positrons signals from the galaxy with the $511\,$keV line~\cite{Laha:2019ssq,Dasgupta:2019cae}, current~\cite{Lee:2021qhe} or prospective~\cite{Ballesteros:2019exr} X-ray limits, neutrino constraints from Super-Kamiokande~\cite{Dasgupta:2019cae}, JUNO~\cite{Wang:2020uvi} or prospective neutrino detectors~\cite{Calabrese:2021zfq,DeRomeri:2021xgy}, gamma ray constraints from INTEGRAL~\cite{Laha:2020ivk}, COMPTEL with improved low-energy secondary particles treatment~\cite{Coogan:2020tuf}, prospective AMEGO instrument~\cite{Coogan:2020tuf,Ray:2021mxu}, LHASSO~\cite{Cai:2021zxo} or fine modelisation of the Galaxy~\cite{Iguaz:2021irx}, prediction of signals from Planet 9 within the PBH hypothesis~\cite{Arbey:2020urq}, archival galactic center radio observations~\cite{Chan:2020zry}, interstellar medium temperature in dwarf galaxies~\cite{Kim:2020ngi,Laha:2020vhg,Lu:2021rnz} or $21\,$cm measurements by EDGES with Schwarzschild~\cite{Mittal:2021egv} or Kerr PBHs~\cite{Natwariya:2021xki}, Big Bang nucleosynthesis (BBN)~\cite{Luo:2020dlg}, heat flow from a small BH captured in the Earth core~\cite{Acevedo:2020gro}, warm DM from light Schwarzschild~\cite{Auffinger:2020afu} and Kerr~\cite{Masina:2021zpu} PBHs, dark radiation from light spinning PBHs~\cite{Masina:2021zpu,Arbey:2021ysg}, (extended) dark sector emission~\cite{Baker:2021btk,Calabrese:2021src}, axion-like particle emission~\cite{Schiavone:2021imu}, HR from extended BH metrics~\cite{Arbey:2021jif,Arbey:2021yke}. While this release note was being written, the authors became aware of Refs.~\cite{Cheek:2021odj,Cheek:2021cfe} which analyze the production of DM by PBH evaporation in the early universe; while these do not explicitly use the code \texttt{BlackHawk}, apart from the spin 2 greybody factors, they provide results compatible with those of \texttt{BlackHawk} with mostly the same spirit in the numerical part.

This paper is a release note of a new version of \texttt{BlackHawk v2.0} that includes some primordial new features linked to the physics described above: dark sector emission, spin $3/2$ greybody factors, BH remnants, BSM BH metrics, low energy hadronization; as well as bug corrections. It then completes the manual published in \textit{European Physics Journal C}~\cite{manual} and goes together with the updated version of the code available on the \texttt{BlackHawk} website mentioned above, and an updated version of the manual available on the arXiv~\cite{manual} (v3). All the installation and run procedures, as well as the complete set of parameters and routines are described in the latter, while we focus here on the new features only. It is organized as follows: in Section~\ref{sec:new_features} we describe the physics of the new features added to the code and illustrate them with examples of \texttt{BlackHawk} results, in Section~\ref{sec:parameters} we list the new input parameters, in Section~\ref{sec:optimisation} we mention some improvements on the optimisation of the code and we conclude in Section~\ref{sec:conclusion}. In the following, we use the natural system of units $G = \hbar = c = k\mrm{B} = 1$.

\section{New features}
\label{sec:new_features}

In this Section, we describe the new features added to \texttt{BlackHawk} since the manual was published~\cite{manual} and we give examples of interesting new results. They have been grouped into distinct categories, even if the changes were made separately: addition of DM, spin $3/2$ greybody factors, time dependent features for cosmological studies, greybody factors from new BH metrics, low energy hadronization. Most of these add-ons imply a modification of some parameters and routines, that are listed respectively in Section~\ref{sec:parameters} and Appendix~\ref{app:routines}.

\subsection{Adding a DM particle}
\label{sec:new_features_DM}

A persistent question of the \texttt{BlackHawk} users was about the possibility of adding a (BSM) particle to the computed SM spectra. The procedure to implement this modification was described in the \texttt{BlackHawk} manual, but the authors found it very convenient to add this feature to the vanilla version of the code. This addition of a new particle was used in warm DM~\cite{Auffinger:2020afu,Masina:2021zpu} and DR~\cite{Arbey:2021ysg} studies where it allowed precision improvement compared to previous analytical approximations.

We need to understand what changes when a particle beyond the SM is Hawking radiated. In fact, as the process is purely gravitational and as the emission of individual particles is independent, any number of additional degrees of freedom can be added without changing the rest of the spectra. The only modification is at the level of the BH lifetime. Indeed, in the integration over the total spectrum to obtain the mass and spin change of the BH, additional dofs provide additional contributions: the mass and spin loss rates are enhanced, thus the BH evaporates faster. When considering only one additional dof, as in~\cite{Auffinger:2020afu,Masina:2021zpu,Arbey:2021ysg}, the change in the lifetime is negligible, but it could be much higher with numerous additional dof as considered in~\cite{Baker:2021btk}. In \texttt{BlackHawk}, this change has been taken into account in alternative $f(M,a^*)$ and $g(M,a^*)$ tables computed by an updated (and simplified) version of the script \texttt{fM.c}. Tables \texttt{fM\_add*.txt}, \texttt{fM\_add*\_nograv.txt}, \texttt{gM\_add*.txt} and \texttt{gM\_add*\_nograv.txt} are read instead of the usual \texttt{fM.txt}, \texttt{fM\_nograv.txt}, \texttt{gM.txt} and \texttt{gM\_nograv.txt}, for the addition of a single massless dof of spin \texttt{*}. The tables $f(M,a^*)$ and $g(M,a^*)$ have further been recomputed with increased precision. The greybody factor tables and fits necessary to recompute the $f,g$ tables are also given in the code source. All these files are available in the folder:
\begin{center}
    \texttt{/scripts/greybody\_scripts/fM}
\end{center}
If the additional particle considered has a rest mass $\mu$ that has to be taken into account, new tables must be computed using the \texttt{fM.c} script to limit the emission of this particle to the usual condition $E>\mu$. Finally, the primary spectrum output is modified by the addition of one column to the file \texttt{instantaneous\_primary\_spectra.txt} (program \texttt{BlackHawk\_inst}) or by a new file \texttt{DM\_primary\_spectrum.txt} (program \texttt{BlackHawk\_tot}). The additional dof is assumed to have no interaction with the SM. Hence, there is no modification of the SM secondary spectra. If it were the case, the user would have to modify the routines computing the secondary spectra to implement any new branching ratio (\textit{e.g.}~DM decay into SM particles). We want to stress that within \texttt{BlackHawk}, the graviton is not considered as a BSM particle and is already treated separately since \texttt{BlackHawk v1.0}.

Tab.~\ref{tab:lifetime} below shows the lifetime reduction when one adds a single massless (DM) dof to the SM. As expected, the relative change is small, and depends on the new particle spin, with a decreased impact as the spin increases for a Schwarzschild BH (spin $a^* = 0$). On the contrary, when the BH is close to Kerr extremal (spin $a^* = 0.9999$), higher spin new particles have a greater impact compared to the Schwarzschild case. The spin hierarchy of the impact of adding one dof is not inverted, as one could expect knowing that higher spin particles have a rate of emission much higher than lower spin particles for a close to extremal Kerr BH. The reason is that as the BH evolves, it loses its spin quite fast and behaves like a Schwarzschild BH for most of its lifetime~\cite{Arbey:2019jmj}, where the low spin particles have the higher impact.

\begin{table}[h!]
    \centering
    \begin{tabular}{c||c|c|c|c|c}
         & $s = 0$ & $s = 1/2$ & $s = 1$ & $s = 3/2$ & $s = 2$ \\
         \hline
         $a^* = 0$ & $1.73\%$ & $0.96\%$ & $0.39\%$ & $0.07\%$ & $0.06\%$ \\
         $a^* = 0.9999$ & $1.43\%$ & $0.93\%$ & $0.51\%$ & $\emptyset$ & $0.70\%$
    \end{tabular}
    \caption{Table of the relative lifetime reduction of a $M\mrm{BH} = 1\,$g BH with spin $a^* = \{0,0.9999\}$, with the addition of a single massless dof to the SM. The greybody factors of a spin $3/2$ particle emitted by a Kerr BH have not yet been computed.}
    \label{tab:lifetime}
\end{table}

\subsection{Spin 3/2 particle}
\label{sec:new_features_32}

\begin{figure}
    \centering
    \includegraphics{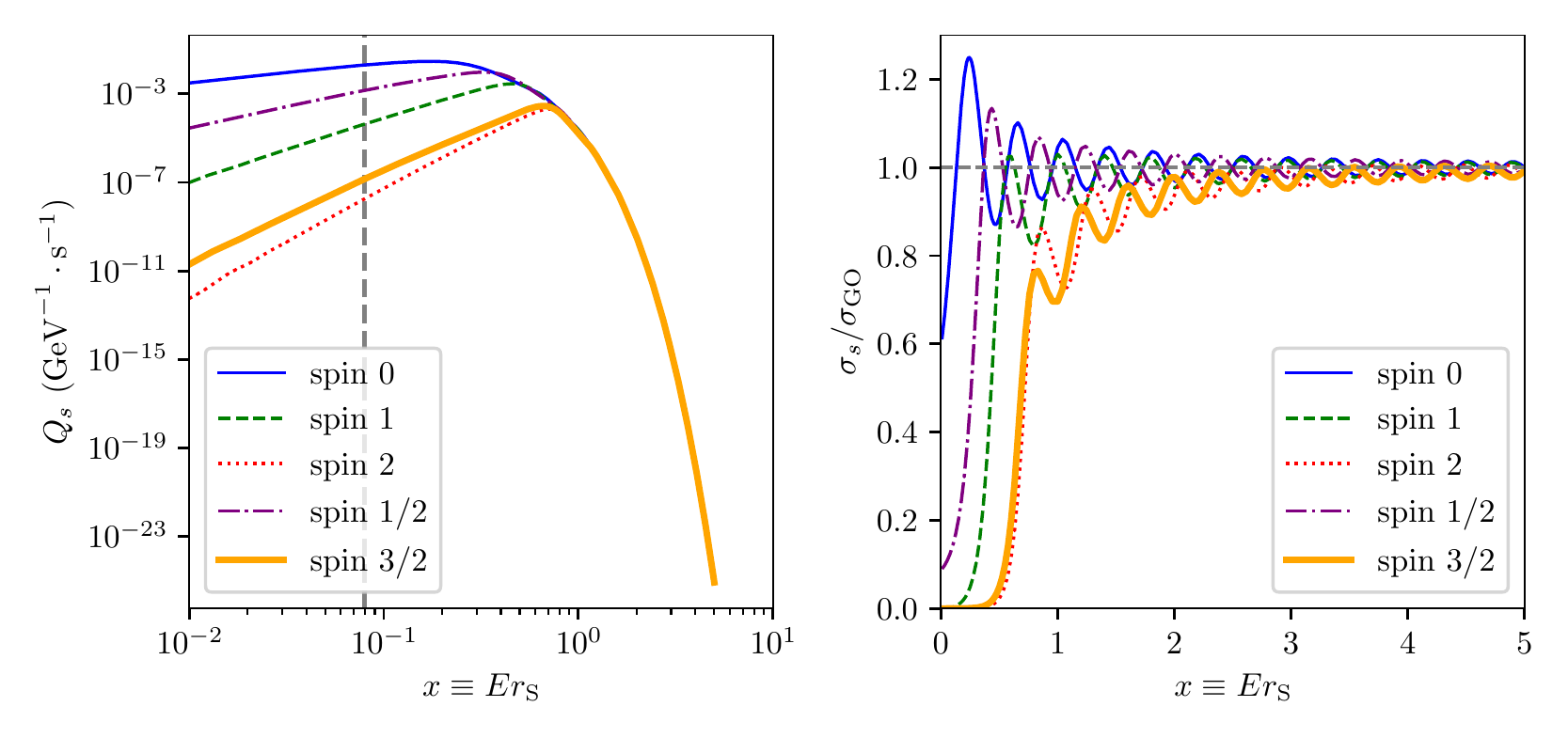}
    \caption{Hawking spectra of massless particles of spins 0 (solid blue), 1 (dashed green), 2 (dotted red) and $1/2$ (dot-dashed purple) as well as the new spectrum of the massless spin $3/2$ fermion (thick solid orange) for a Schwarzschild BH. \textbf{Left:} Full spectra in logarithmic scale, the black vertical line represents the BH temperature. These spectra are shape invariant when changing the BH mass. \textbf{Right:} High energy oscillations of the cross-section, the black horizontal line represents the {\color{blue}geometrical optics} approximation.}
    \label{fig:spin_32}
\end{figure}

When doing DM and DR studies, it became obvious that particles with spin $3/2$ may play a role~\cite{Auffinger:2020afu,Masina:2021zpu}. These particles could be motivated by \textit{e.g.}~supersymmetric extensions of the SM; one example is the gravitino, massive spin $3/2$ superpartner of the graviton.

The Newman--Penrose form of the Rarita-Schwinger equation of motion for the massless spin $3/2$ field in Kerr--Newman geometry, the corresponding radial Teukolsky equation, as well as the short range potential $V_{3/2}(r(r^*))$ obtained following the Chandrasekhar transform of the Teukolsky equation are described in~\cite{TorresDelCastillo:1989hk,TorresdelCastillo:1990aw,TorresdelCastillo:1992zq}. We must then integrate the equation of motion of the spin $3/2$ field from the BH horizon to spatial infinity
\begin{equation}
    \dfrac{\d^2 Z}{\d r^{*2}} + (E^2 - V_{3/2}(r(r^*)))Z = 0\,.\label{eq:wave}
\end{equation}
The greybody factor for some partial spin-weighted spherical harmonic $s,l$ is given as the probability that the emitted particle reaches infinity, expressed in terms of the wave function amplitudes as
\begin{equation}
    \Gamma_s^l = \left|\dfrac{Z^{\rm out}_{\infty}}{Z^{\rm out}_{\rm hor.}}\right|^2,
\end{equation}
where the superscript ``out" refers the the outgoing part of the wave. For the moment, in \texttt{BlackHawk}, we have only implemented the Schwarzschild special case of the spin $3/2$ Hawking radiation. The massive spin $3/2$ particle can be treated as the other spins, by putting an artificial energy cutoff in the Hawking spectrum at the particle rest mass $E > \mu$.

We have computed the greybody factors thanks to a script similar to the ones used for spins 0, 1, 2 and $1/2$ particles, in the Schwarzschild case. The \texttt{Mathematica} script for spin $3/2$ is \texttt{spin\_1.5.m}; it has been added to the \texttt{BlackHawk} folder:
\begin{center}
    \texttt{/scripts/greybody\_scripts/greybody\_factors}
\end{center}
This script performs the integration of Eq.~\eqref{eq:wave} and tabulates the greybody factors summed over the spherical harmonics decomposition of the wave function. For details about this computation see the complete manual~\cite{manual}. We advise to use as a ``spatial infinity" boundary condition at least $r^*_\infty/r\mrm{H} > 300/x$, where $x \equiv Er\mrm{S}$ is the adimensioned energy, for good numerical convergence. This is particularly important for the low energy greybody factors. The script \texttt{fM.c} has also been updated as already developed in Section~\ref{sec:new_features_DM} to include spin $3/2$ emission in the BH lifetime computation.

In Fig.~\ref{fig:spin_32} (left panel) we show the Hawking radiation rate for all massless spins fields with a single dof
\begin{equation}
    Q_s(E,M\mrm{BH}) \equiv \dfrac{\d^2 N_s}{\d t\d E} = \dfrac{\Gamma_s}{e^{E/T\mrm{BH}} - (-1)^{2s}}\,,
\end{equation}
where $s$ is the particle spin, $T = 1/4\pi r\mrm{S}$ is the Schwarzschild BH temperature and
\begin{equation}
    \Gamma_s(E,M\mrm{BH}) \equiv \sum_{l = 0}^{+\infty}(2l+1)\Gamma_s^l(E,M\mrm{BH})\,,
\end{equation}
where $\Gamma_s^l$ is the greybody factor for angular momentum $l$, energy $E$ and BH mass $M\mrm{BH}$. In the same Figure (right panel) we also show the high energy limit of the cross section for all spins
\begin{equation}
    \sigma_s \equiv \dfrac{\pi\Gamma_s}{E^2}\,,
\end{equation}
compared to the geometrical optics (GO) approximation~\cite{MacGibbon:1990zk}
\begin{equation}
    \sigma\mrm{GO} \equiv \dfrac{27}{4}\pi r\mrm{S}^2\,,
\end{equation}
where $r\mrm{S} \equiv 2M$ is the Schwarzschild horizon radius.

\subsection{Time dependent features, cosmological studies}
\label{sec:new_features_time}

\begin{figure}
    \centering
    \includegraphics{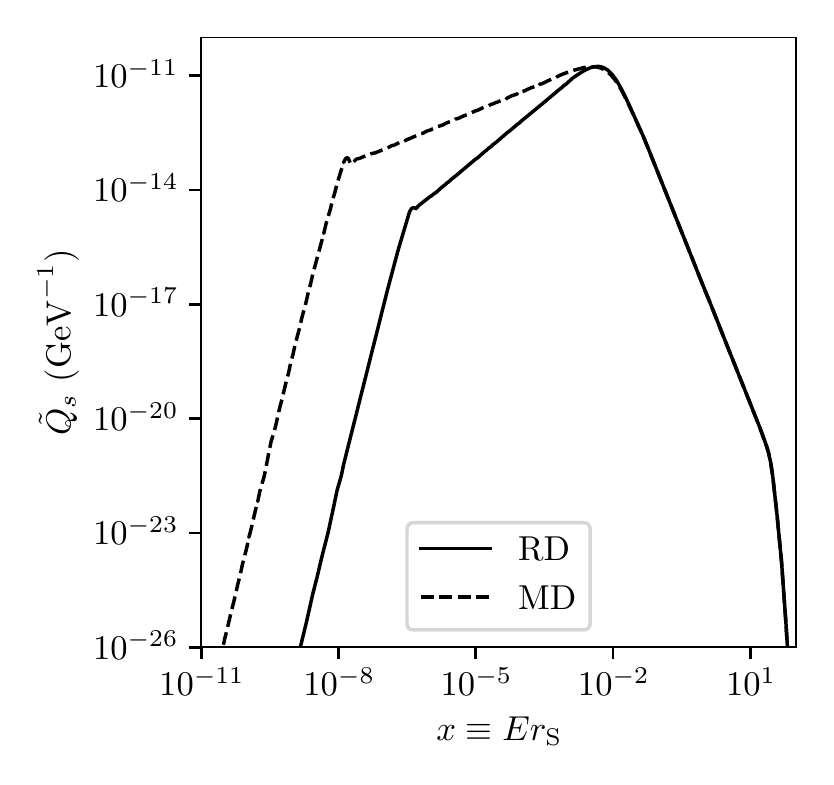}
    \caption{Stacked graviton spectrum emitted by a $M\mrm{BH} = 10^{-1}\,$g Schwarzschild BH, in the case of radiation domination (RD, solid black) or matter domination (MD, dashed black).}
    \label{fig:stacked}
\end{figure}

Several studies performed with \texttt{BlackHawk} consider Hawking radiation through cosmological eras, \textit{e.g.}~the early universe before BBN~\cite{Auffinger:2020afu,Masina:2021zpu,Arbey:2021ysg}. It can be relevant to consider not the full time-dependent spectrum of emitted particles, but only the stacked spectrum of those particles at some time after the BH evaporation. This quantity must be computed by taking the redshift into account. Indeed, particles emitted at the beginning of the BH evaporation see their energy diluted compared to those emitted at the end of the BH lifetime. We thus provide into \texttt{BlackHawk} a \texttt{C} script \texttt{stack.c} that performs this integration, with adjustable cosmology parameters such as the matter/radiation domination at the time of evaporation, the time of matter-radiation equality, \textit{etc}. This script is available in the folder:
\begin{center}
    \texttt{/scripts/cosmology\_scripts}
\end{center}
and is accompanied by a \texttt{README.txt} file containing instructions. This kind of integration requires that we can extract the time steps of the BH evaporation precisely. It was not possible with the previous version of \texttt{BlackHawk} as only the absolute time was available in the output files, with a exponential precision of 5 digits. At the end of the BH evaporation, the time steps become so tiny that this precision is not sufficient to distinguish the time steps, making the integration procedure incomplete. Thus, we have added the output file \texttt{dts.txt} to the result set of a \texttt{BlackHawk} run which contains the explicit time steps on top of the absolute time. This file is read by the script \texttt{stack.c}.

The script computes the integrated spectrum $\tilde{Q}_i$ of particle $i$ from BH formation $t\mrm{form}$ to total evaporation $t\mrm{eva}$
\begin{equation}
    \tilde{Q}_i(E) = \int_{t\mrm{form}}^{t\mrm{eva}} (1+z(t))\dfrac{\d^2 N_i}{\d t \d E}\left(t,(1+z(t))E\right)\,\!\d t\,.\label{eq:stacked}
\end{equation}
The redshift between running time $t$ and the evaporation time $t\mrm{eva}$ is taken into account to shift the energies as well as the BH distribution dilution. This results in the two factors $1+z(t)$ where the redshift as a function of time is given by
\begin{equation}
    1+z\mrm{RD}(t) = \left(\dfrac{t\mrm{eva}}{t}\right)^{1/2},\quad\text{or}\quad 1+z\mrm{MD}(t) = \left(\dfrac{t\mrm{eva}}{t}\right)^{2/3},
\end{equation}
for radiation domination (RD) and matter domination (MD) respectively. The case of a mixed or alternate domination of radiation and matter could be straightforwardly implemented in the routine \texttt{redshift()} inside that script.

On top of that, it has become important to treat precisely the boundaries of integral~\eqref{eq:stacked}. This is why we have implemented in \texttt{BlackHawk} an explicit formula for the value of the BH formation time $t\mrm{form}$ (see \textit{e.g.}~\cite{Carr:2020gox,Auffinger:2020afu})
\begin{equation}
    t\mrm{form} = \dfrac{M\mrm{BH}}{\gamma}\,,\label{eq:time}
\end{equation}
where $\gamma \simeq 0.2$ is the fraction of a Hubble sphere that collapses into a BH when the density perturbation re-enters the Hubble horizon. This formula is valid for BH formation during a radiation dominated era.

In Fig.~\ref{fig:stacked} we show the (massless) graviton spectrum of a Schwarzschild BH stacked over its lifetime considering RD or MD. The shape can be compared to the instantaneous (massless) spin 2 spectrum of Fig.~\ref{fig:spin_32} (left panel).

On a related subject, multiple scenarios of quantum gravity predict that the evaporation of BHs stops at some mass greater than the Planck mass. What is left of the BHs, commonly denoted as BH relics or BH remnants, not to mix up with DM relics, does not evaporate and keeps a constant mass. If the BH was originally spinning or had an electric charge, it is possible that random fluctuations at the end of evaporation attribute a non-zero spin/charge to the remnant~\cite{Page:1977um}. It is also possible that the usual HR paradigm breaks down at some mass scale above the Planck mass, where quantum effects are believed to be important, thus advocating for cautious treatment of the emission below this mass scale. These stable BH remnants, of a mass of the order of magnitude of or superior to the Planck mass, can constitute some fraction of DM~\cite{MacGibbon:1987my,Barrow:1992hq,Dolgov:2000ht,Chen:2002tu,Barrau:2003xp,Lehmann:2019zgt} (see however the new discussion started by~\cite{Kovacik:2021qms}). As the remnant mass depends on the quantum gravity model, and in particular on the number of extra spatial dimensions~\cite{Hossenfelder:2003dy}, we are agnostic about it and leave it as a free parameter of the code. Apart from the particular case of $M\mrm{remnant} \lesssim M\mrm{BH}$, the BH lifetime and stacked spectra should not be perturbed much by the incomplete evaporation.

\subsection{New BH metrics and greybody factors}
\label{sec:metrics}

\begin{figure}
    \centering
    \includegraphics{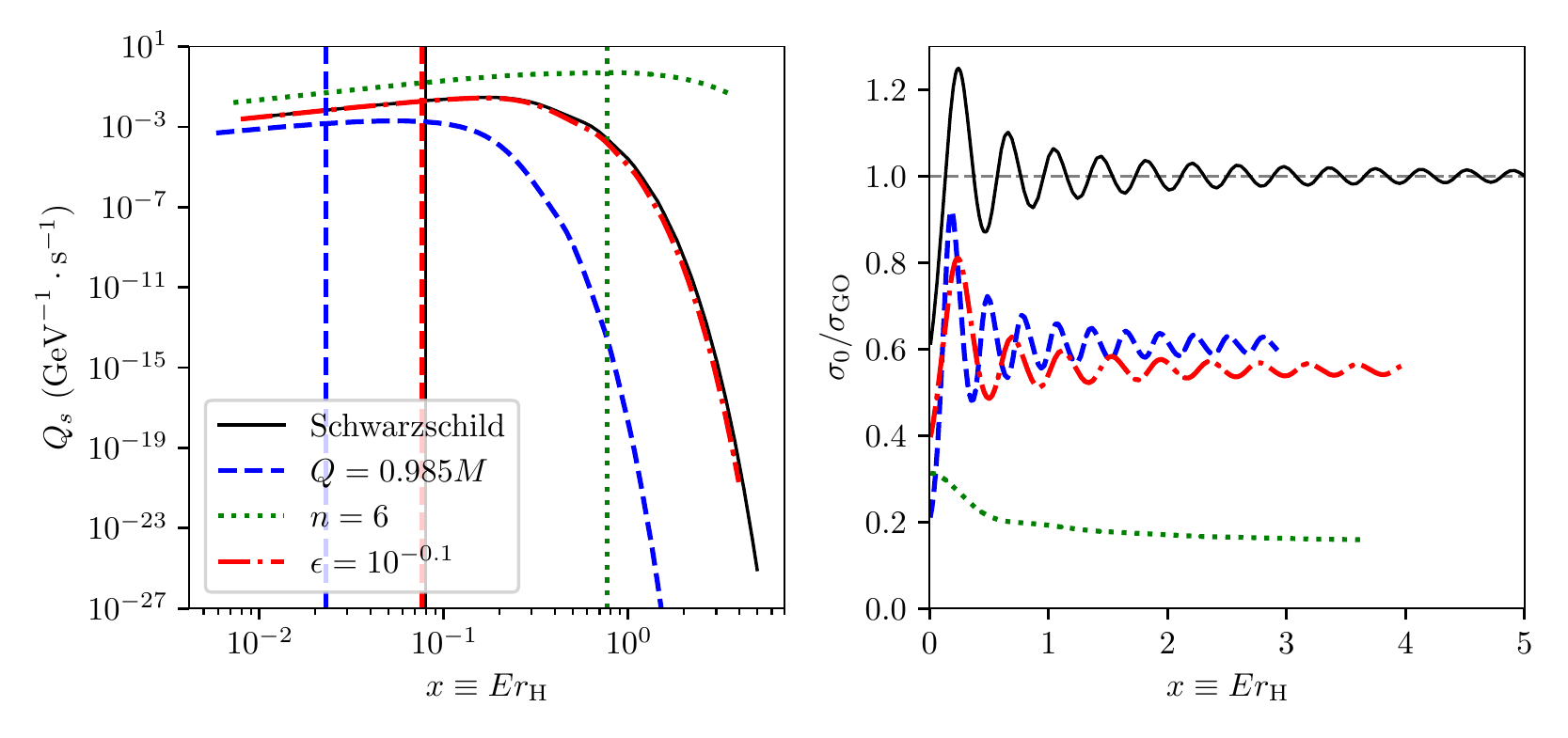}
    \caption{Comparison of the HR of a massless spin 0 particle from a charged BH with charge $Q = 0.985M$ (dashed blue), a higher-dimensional BH with $n = 6$ (dotted green) and a polymerized BH with $\epsilon = 10^{-0.1}$ and $a_0 = 0.11$ (dot-dashed red) to a Schwarzschild BH (solid black) of the same mass $M\mrm{BH}$. The horizontal axis is labelled $x \equiv Er\mrm{H}$ where $r\mrm{H}$ is the BH exterior horizon in the corresponding metric. \textbf{Left:} The full Hawking radiation spectra. The vertical lines denote the BH temperatures, with the same color and style correspondence as the curves. \textbf{Right:} The high energy cross section oscillations. The grey horizontal dashed line denotes the {\color{blue}geometrical optics} approximation.}
    \label{fig:metrics}
\end{figure}

Constraints over the PBH fraction of DM are more and more stringent. Would an experiment detect a signal compatible with PBH Hawking radiation, this would constitute a test of the BH quantum behaviour close to the horizon complementary to the study of BH quantum normal modes during the ringdown phase after a merger (see \emph{e.g.}~\cite{Moulin:2019ekf}). Then, one needs to know precisely what is the effect of exotic BH metrics on the HR signals. In that context, we have implemented greybody factors computed with different BH metrics in \texttt{BlackHawk}, namely higher dimensional BHs, charged (Reissner--Nordstr\"om) BHs and polymerized BHs. The first were already studied in~\cite{Harris:2003eg}, and more recently Ref.~\cite{Johnson:2020tiw} added the higher dimensional greybody factors inside \texttt{BlackHawk} to re-evaluate the extragalactic gamma ray background constraints. Polymerized BHs are one example of a regular BH solution to the Einstein equations that exhibits a horizon with Hawking radiation. Hence, their HR signals, as well as those of other regular BH solutions, are of particular interest and are the subject of numerous recent studies (see \emph{e.g.}~\cite{Rincon:2020cos,Berry:2021hos,Cai:2021ele,Baake:2021jzv,Molina:2021hgx}). Some analytical and numerical results about HR from polymerized BHs were presented in~\cite{Alesci:2011wn,Alesci:2012zz,Hossenfelder:2012tc,Moulin:2018uap,Anacleto:2020zhp}. The literature on charged BHs is older as they are not astrophysically motivated~\cite{Page:1977um}. We completed those results with a full analytical study of the HR by spherically symmetric BHs described by metrics of the type
\begin{equation}
    \d s^2 = -G(r)\d t^2 + \dfrac{1}{F(r)}\d r^2 + H(r)\d\Omega^2\,,
\end{equation}
which are a subset of Petrov type D metrics. In these metrics, the 4-dimensional angular part is $\d\Omega^2 = \d\theta^2 + \sin(\theta)\d\varphi^2$. We derived the short range potentials $V_s$ for the Schr\"odinger-like wave equation equivalent to the radial Teukolsky equation, for massless fields of spins 0, 1, 2 and $1/2$ in Ref.~\cite{Arbey:2021jif}. We presented the Hawking radiation spectra in the companion paper~\cite{Arbey:2021yke}. In the literature, there existed a lot of results for $tr$-symmetric metrics of the common type
\begin{equation}
    F(r) = G(r) \equiv h(r)\,,\quad \text{and} \quad H(r) = r^2\,.
\end{equation}
Examples include higher dimensional BHs (emission on the brane), charged BHs, (A)dS BHs, \textit{etc}. The general case where $F(r) \ne G(r)$ received attention only recently when regular BH solutions of that kind were derived; we computed the greybody factors for polymerized BHs which are one physically motivated example within the loop quantum gravity (LQG) paradigm. The greybody factors for these metrics are computed in the same manner as the Kerr ones in the previous version of the code. The equations of motion of the massless fields of spins 0, 1, 2 and $1/2$ (\textit{i.e.}~the Newman--Penrose equations) are separated into a radial and an angular part. The radial equation can then be transformed into a Schr\"odinger-like wave equation with a short range potential $V_s$ depending on the particle spin and the metrics details
\begin{equation}
    \dfrac{\d^2 Z}{\d r^{*2}} + (E^2 - V_s(r^*))Z = 0\,.
\end{equation}
Solving this equation numerically with planar wave boundary conditions gives access to the reflection and transmission coefficients, the latter being nothing but the greybody factor. Analytical results were obtained in the usual low and high energy limits. For more details about these metrics and their Hawking radiation, see Refs.~\cite{Arbey:2021jif,Arbey:2021yke}.

In Fig.~\ref{fig:metrics}, we show the full Hawking spectra for the massless (uncharged) spin 0 scalar in the case of a charged BH, a higher-dimensional BH, and a polymerized BH (left panel). In the same figure (right panel) we focus on the high energy oscillations of the cross-section.

Inside \texttt{BlackHawk}, these new metrics and their modified greybody factors and $f$ Page factors are available through the \texttt{parameters.txt} file with the new parameter \texttt{metric}. The script \texttt{fM.c} has been modified to include the possibility of computing the function $f(M,\epsilon,a_0)$ for polymerized BHs, for 11 values of $\epsilon \in [ 0,0.79 ]$ and $a_0 = \{0,0.11\}$, and 11 values of $\epsilon \in [1,100]$ and $a_0 = 0$, where the modified Page factor is given by
\begin{equation}
    f(M\mrm{BH},\epsilon) \equiv -M\mrm{BH}^2\dfrac{\d M\mrm{BH}}{\d t} = M^2 \int_0^{+\infty} \dfrac{E}{2\pi}\sum_i \sum_{\rm dof}\dfrac{\Gamma_i(E,M\mrm{BH},\epsilon,a_0)}{e^{E/T\mrm{LQG}} - (-1)^{s_i}}\,\d E\,, 
\end{equation}
where $\Gamma_i(E,M\mrm{BH},\epsilon,a_0)$ is the modified greybody factor for the polymerized metrics with parameters $\epsilon$ and $a_0$. Contrary to the Kerr case, where the BH spin $a^*$ evolves during evaporation, here $\epsilon$ and $a_0$ are purely constant. We want to stress a particular point: if the polymerized BH mass gets too small, namely if $r_+(M,\epsilon)^2 \sim a_0$, then the validity of the solution is yet to be investigated and we do not guarantee the reliability of the Hawking radiation results. The $f$ tables have been computed with and without the graviton contribution and are stored in:
\begin{center}
    \texttt{/src/tables/fM\_tables/fM\_LQG\_*.txt}
\end{center}

On the same model, the greybody factors for the charged BH have been computed for 50 values of $Q \in [0,0.999]$ ($Q$ is taken as positive, without loss of generality as the greybody factors are computed with neutral particles; the difference with charged particles is altogether of some percent~\cite{Page:1977um}) and for the higher dimensional BH for up to 6 extra dimensions $n \in \{0,1,2,3,4,5,6\}$ and $M_* = 1$, with in each case $200$ values of $x \equiv 2EM$ distributed with a new more balanced spread between low and high energies compared to the Kerr case. The \texttt{Mathematica} scripts used to compute them are available in the folders:
\begin{center}
    \texttt{/scripts/greybody\_factors/charged/*.m}\\
    \texttt{/scripts/greybody\_factors/LQG/*.m}\\
    \texttt{/scripts/greybody\_factors/higher/*.m}
\end{center}
accompanied by \texttt{Python} scripts \texttt{exploitation.py} to produce the fitting tables for the high and low energy limits.

It is of course possible to mix the new metrics with DM contribution to the Hawking spectrum (in that case, the Page factor $f$ must be recomputed for polymerized BHs) following the details of Section~\ref{sec:new_features_DM}, at the exception of the spin $3/2$ additional particle, for which the greybody factors have not been computed within the new metrics. Using the precise Hawking radiation rates for primordial black holes, Ref.~\cite{Arbey:2021yke} derived the first ever constraints on polymerized primordial BHs as DM.

\subsection{Low energy photon spectrum}
\label{sec:new_features_hazma}

\begin{figure}
    \centering
    \includegraphics{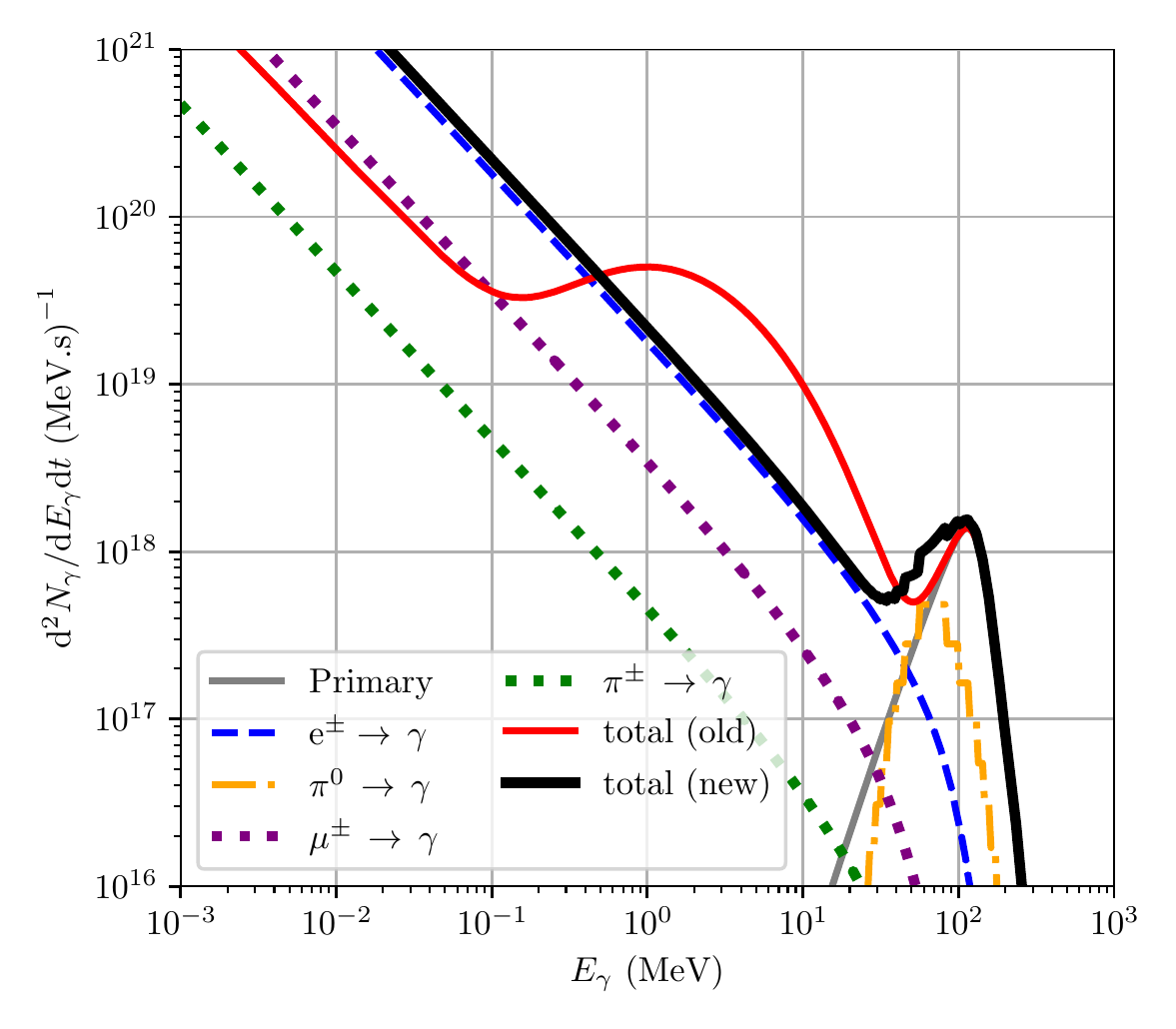}
    \caption{Comparison between the old \texttt{PYTHIA} extrapolated (solid red) and new \texttt{Hazma} computed (solid black) \texttt{BlackHawk} instantaneous total photon spectrum for a $M\mrm{BH} = 5.3\times10^{14}\,$g. From left to right: primary photon spectrum (grey solid); neutral pion decay (orange dot-dashed); electron FSR (blue dashed); muon decay+FSR (purple dotted); charged pion decay+FSR (green dotted). To be compared with Fig.~2 upper left panel of~\cite{Coogan:2020tuf}.}
    \label{fig:hazma_y}
\end{figure}

\begin{figure}
    \centering
    \includegraphics{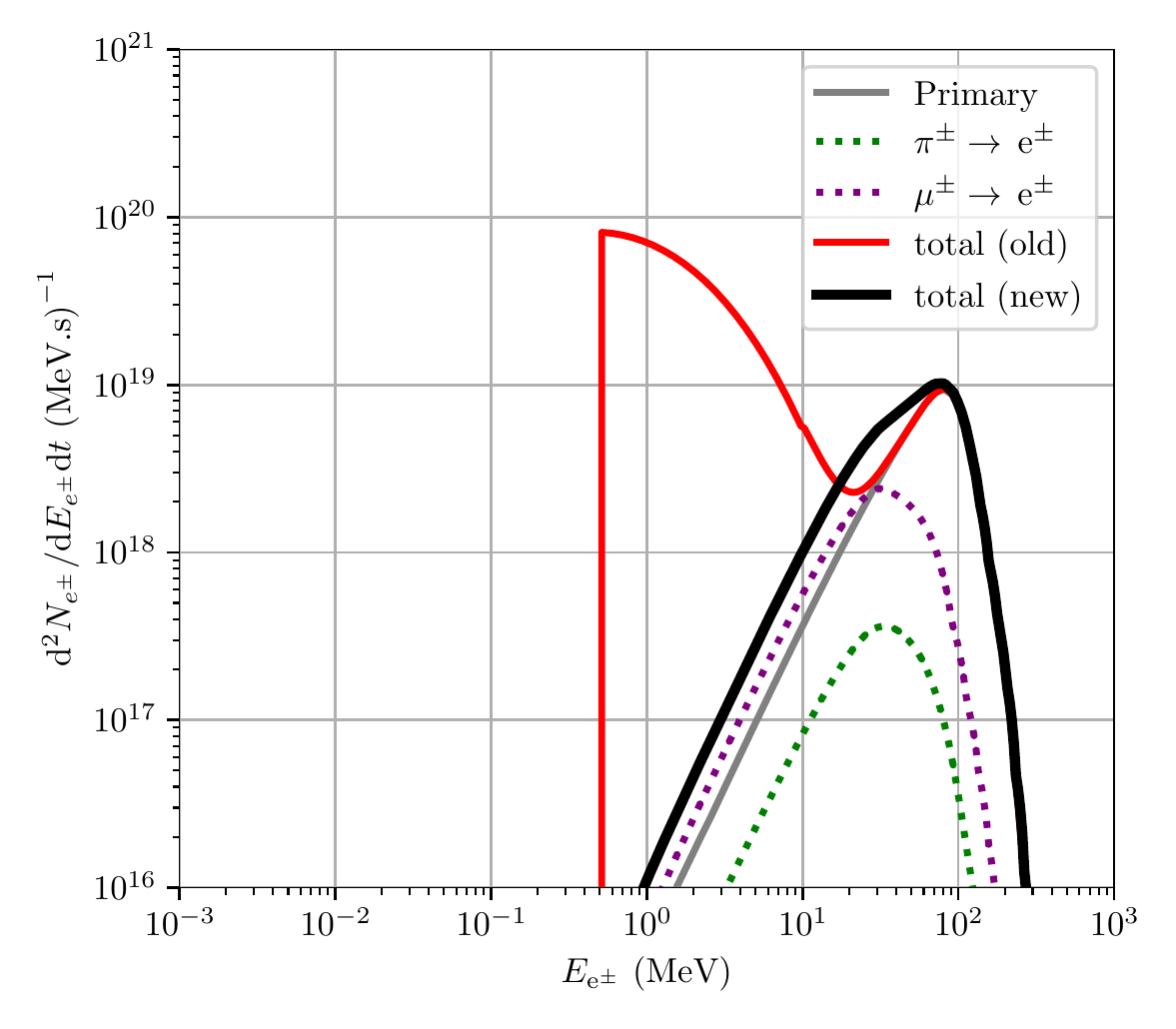}
    \caption{Comparison between the old \texttt{PYTHIA} extrapolated (solid red) and new \texttt{Hazma} computed (solid black) \texttt{BlackHawk} instantaneous total electron spectrum for a $M\mrm{BH} = 5.3\times10^{14}\,$g. From left to right: primary electron spectrum (grey solid); muon decay (purple dotted); charged pion decay (green dotted).}
    \label{fig:hazma_e}
\end{figure}

It was obvious during the development of \texttt{BlackHawk} that secondary spectra would cause a difficulty due to the limits in the theoretical and numerical tools at hand to determine the evolution of SM particles after having been Hawking radiated. In the first version of the code, we used the public particle physics codes \texttt{PYTHIA}~\cite{Sjostrand:2014zea} and \texttt{HERWIG}~\cite{Bellm:2019zci} to convolve the primary spectra with hadronization and decay branching ratios. These codes are designed to match the data of particle accelerators and thus their domain of validity corresponds to the energy range of these accelerators, something like $\sim 5\,$GeV to $\sim10\,$TeV. We decided to extend this range to $100\,$TeV, meaning that there is no unknown physics until there (the fundamental interactions are the same, there is no new (interacting) particle in this range). For higher and lower energies, we decided to extrapolate the branching ratios of \texttt{PYTHIA} and \texttt{HERWIG}, when the secondary particles were kinematically allowed. However this should break down at low energy. Indeed, as stated in~\cite{MacGibbon:1990zk,MacGibbon:1991tj}, the secondary spectra hypothesis is that primary SM particles are emitted at all energies if kinematically allowed ($E > \mu$ where $\mu$ is the particle rest mass), then these fundamental particles are hadronized or decayed. For example, a down quark with a mass of $\simeq 4.7\,$MeV can be emitted at an energy of $E>4.7\,$MeV. But as we know, quarks can not exist outside of bound states, and the lightest bound state is the neutral pion with $\mu_{\pi^0} \simeq 135\,$MeV. The questions is now: \textit{are quarks of type $q$ Hawking emitted between $E = \mu_q$ and $E = \mu_{\pi^0}$?} A question which has been reformulated elsewhere as: \textit{what are the fundamental particles at some low energy $E$?} Indeed, we can argue that below some energy scale (typically the QCD scale), the fundamental degrees of freedom of the theory are the pions and not the individual quarks. Below the pion rest masses, there is no viable QCD degree of freedom, even if light quarks and gluons are kinematically allowed. That is the scheme chosen by~\cite{Coogan:2020tuf} when deriving the secondary photon spectrum at low energy (below the QCD scale).

We thus modified \texttt{BlackHawk} such that it emits pions instead of single quarks if the user chooses so, with the new hadronization choice in the \texttt{hadronization\_choice} parameter. Then, we take back the development of~\cite{Coogan:2020tuf} on the behaviour of the particles at low energy and use the public \texttt{Python} code \texttt{Hazma}~\cite{Coogan:2019qpu} to evolve the primary particles. We restrict ourselves to the secondary photon and electron spectra (\texttt{Hazma} does not provide neutrino spectra), which is an improvement over~\cite{Coogan:2020tuf} which only considered secondary photons. Depending on the available energy, electrons, muons and pions can contribute to the secondary spectra, in addition to the primary photons and electrons. The contributions are:
\begin{itemize}
    \item final state radiation (FSR): pairs of created opposite charge particles radiate photons, this concerns electrons, muons and charged pions;
    \item decays: single unstable particles decay into photons or electrons, this concerns muons, neutral and charged pions.
\end{itemize}
We follow~\cite{Coogan:2020tuf} to compute the FSR photons thanks to the electroweak splitting functions~\cite{Chen:2016wkt} directly inside \texttt{BlackHawk} from the primary electron, muon and charged pions spectra. We also follow~\cite{Coogan:2020tuf} and use the public code \texttt{Hazma}~\cite{Coogan:2019qpu} to compute the low-energy pions and muon decay rates, which we have tabulated in:
\begin{center}
    \texttt{/src/tables/hazma\_tables/photon.txt}
\end{center}
which covers center of mass energy of $1\,$keV to $5\,$GeV. We improve on~\cite{Coogan:2020tuf} by including the secondary low energy electron and positron spectra, from the decay rates of muons and charged pions tabulated in:
\begin{center}
    \texttt{/src/tables/hazma\_tables/electron.txt}
\end{center}
The \texttt{Python} script we have used to compute these tables is available at:
\begin{center}
    \texttt{/scripts/hazma\_scripts/hazma\_tables.py}
\end{center}
Then, we add up all the contributions to the primary photon and electron spectra to obtain the total secondary spectra.

There are several drawbacks to this methodology. First, when FSR occurs, there is a loss of center of mass energy into photons, which should alter the charged particles spectra before decays, even if at second order in the fine structure constant. That difficulty arises because the \texttt{Hazma} code convolves initial spectra with analytical transfer functions and does not perform a MCMC statistical analysis of competing processes. Second, and this is the main difficulty with the QCD phenomenology, it is not clear what is precisely the QCD scale $\Lambda\mrm{QCD}$ (between $\mu_{\pi^0}$ and $300\,$MeV) that separates direct pion emission from single quarks and gluons emission that hadronize into jets. This is the reason why we do not fix a $\Lambda\mrm{QCD}$ at some value, but rather provide both hadronization possibilities. The spectra should interpolate between the two limits when the center of mass energy explores the QCD range. As we can not decide (outside of a chosen model) what are the number of degrees of freedom and their nature in the QCD energy range, this hadronization method is limited to the program \texttt{BlackHawk\_inst}, because we need this information to compute the Page evolution factors inside \texttt{BlackHawk\_tot}. The PBH masses at stake are $M\mrm{PBH}\gtrsim 10^{14}\,$g, thus we can consider that these PBHs are living indefinitely at our time scale --- the age of the universe --- and that \texttt{BlackHawk\_inst} is enough for PBH studies linked to photon or electron emission in this scheme.

We show in Fig.~\ref{fig:hazma_y} (respectively Fig.~\ref{fig:hazma_e}) the total photon (respectively electron) spectrum computed with the old extrapolated version of \texttt{BlackHawk}, compared to the new spectra computed with the method of~\cite{Coogan:2020tuf} for a BH of mass $5.3\times10^{14}\,$g.

We point out that this new feature is dedicated to the low energy photon and electron emission, and does not for now provide low energy spectra for neutrinos, which however are only related to sub-dominant (but complementary) constraints in the considered PBH mass range~\cite{Dasgupta:2019cae,Wang:2020uvi,Calabrese:2021zfq,DeRomeri:2021xgy}. We stress that the new spectra may alter the constraints of previous gamma ray and electron-positron studies from Hawking evaporation in the narrow mass range $5\times 10^{14}\,{\rm g}\lesssim M\mrm{PBH} \lesssim 10^{16}\,$g, where PBHs can only represent a negligible fraction of DM anyway~\cite{Green:2020jor,Carr:2020gox}.

\section{Parameters}
\label{sec:parameters}

In this Section we describe the new \texttt{parameters.txt} file resulting from the modifications and add-ons described in Section~\ref{sec:new_features}, as well as a simplification of the presentation of the file by a removal of ``hard-coded" parameters such as the dimensions of the numerical tables. These parameters are read by new routines in new \texttt{infos.txt} files. On another hand, as we have added new metrics, some parameters changed their name from \texttt{*\_a} (Kerr case) to \texttt{*\_param} in order to be generalized. As a consequence, the extended distributions for the BH spin $a^*$ have been adapted to the BH charge $Q$, while for polymerized and higher dimensional BHs there is only one value of the secondary parameter allowed for each run: (\texttt{a\_number} $\rightarrow$) \texttt{param\_number = 1}. We focus here only in the new or modified parameters which are the interesting entries for \texttt{BlackHawk} users. For a complete description of the new and modified routines see Appendix~\ref{app:routines} in this paper or the complete updated \texttt{BlackHawk v2.0} manual~\cite{manual}. Some bug corrections are listed in Appendix~\ref{app:corrections}. The parameters file/structure have several interesting new entries:
\begin{itemize}
    \item the new parameter \texttt{metric} switches between the Kerr metric (0), the polymerized metric (1), the Reissner--Nordstr\"om metric (2) and the higher-dimensional metric (3);
    \item for the new BH metrics, we have added the parameters \texttt{Qmin} and \texttt{Qmax} (dimensionless --- positive --- BH charge between $0$ and $1$), \texttt{epsilon\_LQG} and \texttt{a0\_LQG} (polymerized metric, $\epsilon$ being the polymerization parameters between 0 and 1 and $a_0$ the minimal dimensionless area), and finally \texttt{n} and \texttt{M\_star} (higher dimensional metric, $n$ is the number of extra dimensions and $M_*$ the rescaled Planck mass);
    \item \texttt{tmin\_manual}: as described in Section~\ref{sec:new_features_time}, there is now a possibility to choose between a manual BH formation time $t\mrm{form}$ (\texttt{tmin\_manual = 1}, \texttt{tmin} $= t\mrm{form}$) and the fidutial $t\mrm{form}$ given by Eq.~\eqref{eq:time} (\texttt{tmin\_manual = 0}), the latter being only relevant for monochromatic BH distributions and radiation domination at formation;
    \item \texttt{nb\_final\_times} has been removed as it is now fixed by the integration procedure;
    \item \texttt{nb\_gamma\_spins} describes how many particle spins are available for some BH metric, and \texttt{nb\_gamma\_fits} encodes the number of fitting parameters used to extend the greybody factor tables to low and high energy;
    \item \texttt{add\_DM}, \texttt{m\_DM}, \texttt{spin\_DM} and \texttt{dof\_DM} are new parameters linked to the DM emission described in Section~\ref{sec:new_features_DM}. To add DM emission one has to set \texttt{add\_DM = 1} (otherwise let it to 0), then one can choose the DM mass $m\mrm{DM}$ in GeV, the DM spin $s\mrm{DM} \in\{ 0, 1, 2, 0.5, 1.5\}$ and the DM number of dof (be carefull that only 1 dof has been taken into account when computing the new $f,g$ tables with DM emission);
    \item \texttt{BH\_remnant} switches between the usual total evaporation at Planck mass (\texttt{BH\_remnant = 0}) and an interrupted evaporation at \texttt{M\_remnant} $\ge M\mrm{P}$ (\texttt{BH\_remnant = 1});
    \item the parameter \texttt{hadronization\_choice} can now take the value 3, corresponding to the hadronization procedure described in Sec.~\ref{sec:new_features_hazma}. This choice forces the DM parameters to be used to compute the pion spectrum, thus \textit{the low energy hadronization procedure is at present incompatible with DM emission by BHs};
    
    \item the Page factors tables parameters (\texttt{Mmin\_fM}, \texttt{Mmax\_fM}, \texttt{nb\_fM\_masses} and \texttt{nb\_fM\_a} $\rightarrow$ \texttt{nb\_fM\_param}), the greybody factor tables parameters (\texttt{particle\_number}, \texttt{nb\_gamma\_a} $\rightarrow$ \texttt{nb\_gamma\_param}, \texttt{nb\_gamma\_x}, \texttt{a\_min} $\rightarrow$ \texttt{param\_min} and \texttt{a\_max} $\rightarrow$ \texttt{param\_max}) and the hadronization parameters (\texttt{Emin\_hadro}, \texttt{Emax\_hadro}, \texttt{nb\_init\_en}, \texttt{nb\_fin\_en}, \texttt{nb\_init\_part} and \texttt{nb\_fin\_part}) do not appear anymore in the \texttt{parameters.txt} file as they are \textit{a priori} fixed, they are now read in info files stored in respectively:
\begin{center}
    \texttt{/src/tables/fM\_tables/infos.txt}
    
    \texttt{/src/tables/gamma\_tables/infos.txt}
    
    \texttt{/src/tables/herwig\_tables/infos.txt}
    
    \texttt{/src/tables/pythia\_tables/infos.txt}
    
    \texttt{/src/tables/pythia\_tables\_new/infos.txt}
    
    \texttt{/src/tables/hazma\_tables/infos.txt}
\end{center}
    and we recommend that they be not modified unless the tables are recomputed;
    \item the following parameters changed their name: \texttt{BHnumber} $\rightarrow$ \texttt{BH\_number}, \texttt{Enumber} $\rightarrow$ \texttt{E\_number}, \texttt{anumber} $\rightarrow$ \texttt{param\_number}.
\end{itemize}

\section{Optimisation}
\label{sec:optimisation}

In this Section we briefly present a simple optimisation procedure that has been implemented inside \BH in order to fasten the computations and diminish the disc memory occupied by the output files when some specific studies do not necessitate the full Hawking radiation spectra. Following the addition of the \texttt{write\_*} parameters which decide whether some particle spectrum is written down to the output, we added the tables \texttt{compute\_primary[]} and \texttt{compute\_secondary[]} which allow the code to skip some particle spectrum computations. We must warn the user that if some primary particle ``computation" parameter is set to 0, then this particle will not participate in the secondary spectra as well. However, for studies focused on a single particle of some type, \textit{e.g.}~the graviton, DM or some secondary particle, then the other secondary particles can be ``turned off". The BH time evolution will not be affected as it is determined by the numerical tables of the Page factors $f,g$.

\section{Conclusion}
\label{sec:conclusion}

We have described the new features available in the public code \texttt{BlackHawk v2.0}, together with interesting illustrations: the addition of dark matter (and dark radiation) emission, the massless spin $3/2$ field greybody factors for the Schwarzschild metric, the possibility to keep a black hole remnant at the end of evaporation, the greybody factors associated with spherically symmetric and static black hole solutions more general than the Schwarzschild case and the careful computation of the low energy photon and electron spectra. We hope that this will open a new era of Hawking radiation studies with already promising results. An updated version of the complete manual is available on the arXiv~\cite{manual} (v3) and the \texttt{BlackHawk v2.0} code is available on HEPForge at \url{https://blackhawk.hepforge.org/}.

\newpage

%%%%%%%%%%%%%%%%%%%%%%%%%%%% APPENDIX %%%%%%%%%%%%%%%%%%%%%%%%%%%%%%%%%%%%
\appendix

\section{Routines}
\label{app:routines}

In this Section we list the new and modified routines of the two programs \texttt{BlackHawk\_inst} and \texttt{BlackHawk\_tot}, resulting from the new features described in Section~\ref{sec:new_features}, as well as the modified parameters listed in Section~\ref{sec:parameters}. We do not list the modifications to the header file \texttt{include.h} and compilation file \texttt{Makefile} as they are transparent regarding the content of Section~\ref{sec:parameters} and of the present Section.

\subsection{Modified routines}

The modifications of existing routines are:
\begin{itemize}
    \item \texttt{int read\_params(struct param *parameters, char name[], int session)} has been modified to read the new parameters described in Section~\ref{sec:parameters} and received bug corrections when parameters were in conflict, as well as a new error display;
    \item \texttt{int memory\_estimation(struct param *parameters, int session)} has been updated to take into account the new \BH features (\textit{e.g.}~the addition of DM to the number of particles simulated);
    \item \texttt{void read\_gamma\_tables(double ***gammas, double *gamma\_a, double *gamma\_x, struct param\\ *parameters)} and \texttt{void read\_asymp\_fits(double ***fits,struct param *parameters)} now include the spin $3/2$ greybody factor tables;
    \item \texttt{void instantaneous\_primary\_spectrum(double **instantaneous\_primary\_spectra, double **BH\_masses, double **BH\_spins, double **spec\_table, double *energies,
	double ***gammas, double *gamma\_a, double *gamma\_x, double ***fits, double *dof, double *spins, double *masses\_primary, int\\ **counters\_a, int ***counters\_x,	int *compute, struct param *parameters)} now takes as arguments the tables \texttt{counters\_x[]}, \texttt{counters\_a[]} and \texttt{compute[]} (for the latter see Section~\ref{sec:optimisation}) to avoid re-definition of those tables at each time step in the program \texttt{BlackHawk\_tot};
	\item \texttt{void write\_instantaneous\_primary\_spectra(double **instantaneous\_primary\_spectra, double\\ *energies, struct param *parameters)}, \texttt{void write\_instantaneous\_secondary\_spectra(double\\ **instantaneous\_integrated\_hadronized\_spectra, double *hadronized\_energies, struct param\\ *parameters)} and \texttt{void write\_lines(char **file\_names\_primary, char **file\_names\_secondary, double **partial\_primary\_spectra, double **partial\_integrated\_hadronized\_spectra,
	int *write\_primary, int *write\_secondary, double time, struct param *parameters)} write down the spectrum of DM if \texttt{add\_DM = 1};
	\item \texttt{void total\_spectra(double ***partial\_hadronized\_spectra, double **partial\_primary\_spectra, double **partial\_integrated\_hadronized\_spectra,
	double ****tables, double *initial\_energies, double *final\_energies, double **spec\_table, double *times, double ***life\_masses, double **BH\_masses, double ***life\_spins, double **BH\_spins, double *energies, double *masses\_primary, double *spins, double *dof, double *masses\_secondary,	double ***gammas, double *gamma\_a, double *gamma\_x, double ***fits, struct param *parameters)} now takes into account the computation of the DM spectrum if \texttt{add\_DM = 1}, it takes the tables \texttt{masses\_primary[]}, \texttt{spins[]} and \texttt{dof[]} as arguments to homogenize the structure of the code and defines the tables \texttt{compute\_primary[]} and \texttt{compute\_secondary[]} (see Section~\ref{sec:optimisation});
	\item \texttt{void life\_evolution(double ***life\_masses, double ***life\_spins, double *life\_times, double *dts, double *init\_masses,
	double *init\_spins, int **rank, double **fM\_table, double **gM\_table, double *fM\_masses, double *fM\_a, struct param *parameters)} fills a table \texttt{dts[]} containing the time steps of integration that are useful for precise time dependent studies (see Section~\ref{sec:new_features_time});
	\item the routines \texttt{void read\_hadronization\_infos(struct param *parameters)}, \texttt{void \\ read\_hadronization\_tables(double ****tables, double *initial\_energies, double *final\_energies, struct param *parameters)}, \texttt{double contribution\_instantaneous(int j, int counter, int k, double **instantaneous\_primary\_spectra, double ****tables, double *initial\_energies, double\\ *final\_energies, int particle\_type, struct param *parameters)} and \texttt{void hadronize\_instantaneous(double ***instantaneous\_hadronized\_spectra, double ****tables, double *initial\_energies, double\\ *final\_energies, double **instantaneous\_primary\_spectra, double *energies, double *masses\_secondary, int *compute, struct param *parameters)} have been modified to read and use the new hadronization tables computed with \texttt{Hazma} (see Sec.~\ref{sec:new_features_hazma}), while \texttt{void convert\_hadronization\_tables(double ****tables, double *initial\_energies, double *final\_energies, struct param *parameters)} has been extended and used to generate the new header file:
\begin{center}
    \texttt{/src/hadro\_hazma.h}
\end{center}
    compiled within the \texttt{BlackHawk} programs if \texttt{HARDTABLES} is defined.
\end{itemize}

We have modified the evolution routines in order to account for the new BH metrics, associated with so-called BH ``secondary parameters" (the BH charge, the value of $\epsilon$ for polymerized BHs or the number of extra dimensions). In fact, all these parameters behave in the same manner as the spin. Some can even be treated as constants, like the number of extra dimensions.

Most routines also take into account the fact that the number of primary particles is now given by:
\begin{center}
    \texttt{particle\_number + grav + add\_DM}
\end{center}

\subsection{New routines}

The new routines are:
\begin{itemize}
    \item \texttt{int read\_fM\_infos(struct param *parameters)} reads the $f,g$ tables info in the new file:
\begin{center}
    \texttt{/src/tables/fM\_tables/infos.txt}
\end{center}
and fills the parameters \texttt{Mmin\_fM}, \texttt{Mmax\_fM}, \texttt{nb\_fM\_masses} and \texttt{nb\_fM\_param};
    \item \texttt{int read\_gamma\_infos(struct param *parameters)} reads the greybody factors tables info in the new files:
\begin{center}
    \texttt{/src/tables/gamma\_tables/infos.txt}
\end{center}
and fills the parameters \texttt{particle\_number}, \texttt{nb\_gamma\_param}, \texttt{nb\_gamma\_x}, \texttt{param\_min} and \texttt{param\_max};
    \item \texttt{int read\_hadronization\_infos(struct param *parameters)} reads the hadronization tables info in one of the new files:
\begin{center}
    \texttt{/src/tables/herwig\_tables/infos.txt}\\
    \texttt{/src/tables/pythia\_tables/infos.txt}\\
    \texttt{/src/tables/pythia\_tables\_new/infos.txt}\\
    \texttt{/src/tables/hazma\_tables/infos.txt}
\end{center}
depending on the \texttt{hadronization\_choice} parameter;
    \item the routine \texttt{void write\_life\_evolutions(double ***life\_masses, double ***life\_spins, double *life\_times, double *dts, struct param *parameters)} writes down the new output file \texttt{dts.txt} (see Sec.~\ref{sec:new_features_time});
    \item new temperature functions \texttt{double temp\_Kerr(double M, double a)}, \texttt{double temp\_LQG(double M, double epsilon, double a0)}, \texttt{double temp\_charged(double M, double Q)} and \texttt{double temp\_higher(double M, double n, double M\_star)} have been written to compute the Hawking temperature for the new BH metrics, as well as some secondary necessary functions \texttt{double rplus\_Kerr(double M, double a)}, \texttt{double P\_LQG(double epsilon)}, \texttt{double m\_LQG(double M, double epsilon)}, \texttt{double rplus\_charged(double M, double Q)}, \texttt{double rminus\_charged(double M, double Q)} and \texttt{rH\_higher(double M, double n, double M\_star)};
    \item \texttt{void add\_FSR\_instantaneous(double **instantaneous\_primary\_spectra, double\\ **instantaneous\_integrated\_hadronized\_spectra,
	double *energies, double *final\_energies, double *masses\_primary, struct param *parameters)} computes the FSR of electrons, muons and charged pions as described in Sec.~\ref{sec:new_features_hazma}; we have been careful about the number of degrees of freedom, as for example the ``electron" spectrum in \texttt{BlackHawk} in fact takes into account the e$^\pm$ multiplicity, we have to divide the formulas of~\cite{Coogan:2019qpu} by 2 because they also account for it a second time;
    \item new \texttt{void free\_*()} functions have been defined to free arrays of different formats;
    \item \texttt{double exp\_adapt(double x)} computes the quantity $\exp(x) - 1$ in the case of small $x$, using the Taylor development of the exponential function up to 5th order.
\end{itemize}

\section{Corrections}
\label{app:corrections}

In this Section, we briefly mention the few typo corrections that were spotted in the manual~\cite{manual} when developing this new version of the code:
\begin{itemize}
    \item There was a double mistake in the definition of $K$ below Eqs.~(2.19) and (2.22). It should be $K \equiv r^2 E$ and $K \equiv (r^2 + a^2)E + am$ respectively, as shows dimensional analysis.
    \item The results presented in Appendix~D are computed for $M\mrm{BH} = 10^{10}\,$g, as shows \textit{e.g.}~the density in Appendix~D.2, whereas the file \texttt{parameters.txt} mentions $M\mrm{BH} = 10^{15}\,$g in Appendix~D.1.
    \item Between Eqs.~(2.13) and (2.14), it is said that the sum over the angular momentum modes $m = -l,...,l$, when all of them give the same contribution to the greybody factor, results in a factor $l(l+1)$. It is of course a factor $2l+1$.
    \item As there is an ambiguity on the precise definition of the BH mass distribution denoted as ``log-normal", we have decided to add both a ``log-normal" distribution for the mass and for the number density of BHs (\texttt{spectrum\_choice = 1} and \texttt{spectrum\_choice = 11} respectively).
\end{itemize}

\bibliography{biblio}

\end{document}